\documentclass[twocolumn,showpacs,prl,amsmath,amssymb,floatfix]{revtex4}
\usepackage{subfiles}
\usepackage{graphicx}
\usepackage{color}

\newcommand{\eqn}[1]    {(\ref{#1})}

\def\bi         {\begin{itemize}}
\def\ei         {\end{itemize}}
\def\benu	{\begin{enumerate}}
\def\eenu	{\end{enumerate}}
\def\bmat       {\left[ \begin{array}}
\def\emat       {\end{array} \right]}
\def\beq	{\begin{equation}}
\def\eeq	{\end{equation}}
\def\beqn       {\begin{eqnarray*}}
\def\eeqn       {\end{eqnarray*}}
\def\beqa       {\begin{eqnarray}}
\def\eeqa       {\end{eqnarray}}
\def\bquote	{\begin{quote}}
\def\equote	{\end{quote}}

\def\bwide	{\begin{widetext}}
\def\ewide	{\end{widetext}}

\def\m          {\mu}

\def\bk         {{\bf k}}

\def\dag	{\dagger}

\begin{document}


\title{Dirac-semimetal phase diagram of two-dimensional black
phosphorus}

\author{Hyeonjin Doh}
\email[Email: ]{clotho72@yonsei.ac.kr}

\author{Hyoung Joon Choi}
\email[Email: ]{h.j.choi@yonsei.ac.kr}
\affiliation{Department of Physics and
Center for Computational Studies of Advanced Electronic Material Properties,
Yonsei University, Seoul 03722, Korea}


\date{\today}

\begin{abstract}
Black phosphorus (BP), a layered van der Waals material, reportedly has 
a band gap sensitive to external perturbations and manifests a Dirac-semimetal phase when its band gap is closed.
Previous studies were focused on effects of each perturbation,
lacking a unified picture for the band-gap closing and the Dirac-semimetal phase.
Here, using pseudospins from the glide-reflection symmetry, we study the electronic structures of mono- and bilayer BP and construct
the phase diagram of the Dirac-semimetal phase in the parameter space
related to pressure, strain, and electric field.
We find that the Dirac-semimetal phase in BP layers is singly connected in the phase diagram, indicating
the phase is topologically identical regardless 
of the gap-closing mechanism. Our findings can be generalized to 
the Dirac semimetal phase in anisotropic layered materials and can play a guiding role in
search for a new class of topological materials and devices.
\end{abstract}


\maketitle


Two-dimensional (2D) materials have attracted much attention in
applications and theories, since graphene was found to be easily producible 
through mechanical exfoliation. Graphene now becomes the typical example 
for the 2D Dirac semimetal with the linear band structure and high mobility. 
However, without a band gap, graphene has a limitation in applications 
for optical or switching devices. On the other hand, another group of 
2D materials, such as transition-metal dichalchogenides, have 
intermediate band gaps and show applicability to electronic devices. Recently, 
black phosphorus (BP) \cite{Bridgman1914}, an allotrope of phosphorus,
gains lots of attention as a new 2D layered 
system \cite{Li2014,Xia2014,Liu2014}.
BP has a thickness-dependent band gap varying from 0.3 eV for the 
bulk \cite{Keyes1953,Takao1981,Akahama1983,Morita1986} up to $\sim 2$ eV 
for the monolayer \cite{Takao1981,Liang2014,Tran2014,Li2016}, and it is 
remarkable with the emergence of the Dirac-semimetal 
phase \cite{Kim2015,Baik2015,Liu2015,Fei2015,Dolui2015,Xiang2015,Gong2016,Ezawa2014,Dutreix2016,Yuan2016}
and the quantum Hall effect at low temperature \cite{Li2016a}.

BP has a layered puckered-honeycomb structure, as shown in 
Fig.~\ref{fig:BPbiStruct} \cite{Hultgren1935,Brown1965,Cartz1979}.
The interlayer interaction is mainly van der Waals 
interaction \cite{Pauling1952,Appalakondaiah2012}, which makes the bulk 
sample easily cleaved to few-layered 2D samples. 
BP layers have strong in-plane anisotropy in the atomic structure, resulting
in anisotropic electrical 
conduction \cite{Akahama1983,Morita1986,Liu2014,Fei2014,Rudenko2016,Low2015}. 
The band gap of BP is controllable by various external parameters such as 
pressure \cite{Keyes1953,Takao1981,Morita1986,Xiang2015,Gong2016,Rodin2014},
electric field \cite{Guo2014,Dai2014,Liu2015,Dolui2015},
and potassium doping \cite{Kim2015,Baik2015}.
Very recently, BP is found to become a Dirac semimetal when its gap is 
closed \cite{Kim2015,Baik2015,Liu2015,Fei2015,Dolui2015,Xiang2015,Gong2016},
making a strong contrast with other 2D semiconducting materials.
Electronic structures of BP have been studied successfully
by tight-binding methods \cite{Takao1981,Rudenko2014,Rudenko2015},
density-functional 
methods \cite{Kim2015,Baik2015,Liu2015,Fei2015,Dolui2015,Gong2016,Appalakondaiah2012,Fei2014,Rodin2014,Guo2014,Rudenko2014,Rudenko2015},
and the GW method \cite{Rudenko2014,Rudenko2015}.
However, a unified picture for the band-gap closing method and
the Dirac-semimetal phase in BP layers is still lacking.

\begin{figure}
\includegraphics[width=7cm]{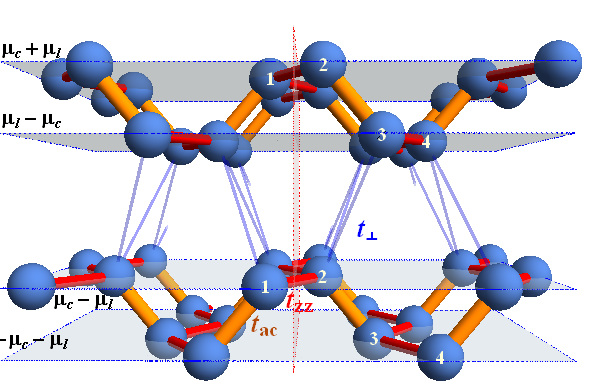}
\caption{\label{fig:BPbiStruct}
(Color online) Atomic structure of bilayer BP.
Red, orange, and blue bonds denote the intralayer hopping energy 
along the zigzag and armchair directions ($t_{zz}$ and $t_{ac}$)
and the interlayer hopping energy ($t_\perp$), respectively.
The vertical red dotted plane denotes the glide-reflection symmetric plane.
Each horizontal plane may have a different potential energy
due to a vertical electric field. 
$\mu_c$ and $\mu_l$ are intra- and interlayer potential
difference, respectively. The number at each phosphorus
atom is for the basis index in the unit cell.
}
\end{figure}

In this letter, we investigate key ingredients of BP to obtain a unified 
picture for the band-gap closing and the Dirac-semimetal phase. 
The gap-closing condition is obtained in the parameter space of 
Hamiltonian that is related to pressure, strain, and external electric fields. 
From the glide-reflection (GR) symmetry of BP, pseudospins are derived, 
which guarantee a massless Dirac Hamiltonian when the band gap is closed. 
In each layer, the band gap can be closed by reducing the anisotropy 
of the intralayer hopping, while the interlayer coupling narrows 
the band gap in multilayer BP. We also find intra- and interlayer 
potential differences have opposite effects so that a 
perpendicular electric field does not decrease the band gap in the monolayer 
while it reduces the gap in the multilayer. Obtained phase diagrams show 
that the Dirac semimetal phase in BP layers is topologically identical 
regardless of the gap-closing method despite some difference in detailed 
electronic band structures.

Layered BP has four P atoms in the unit cell, as indexed in 
Fig.~\ref{fig:BPbiStruct}.
For a tight-binding model, one orbital per P atom can
describe the electronic structure near the band gap \cite{Rudenko2015}.
Based on the tight-binding parameters extracted from
the GW calculation \cite{Rudenko2015},
we construct a simple Hamiltonian which includes
the nearest-neighbor intralayer hopping energies, 
$t_{zz} = -1.486$~eV and $t_{ac} = 3.729$~eV,
along the zigzag and armchair directions, respectively,
and the nearest-neighbor interlayer hopping energy $t_\perp = 0.524$~eV.
Although we use these parameters to describe pristine BP systems,
our major findings are valid independently of their detailed values.
For the monolayer, the Hamiltonian in the momentum space is
$H=\sum_{\bf k}\Psi_{\bf k}^\dag
\hat{H_{\bf k}} \Psi_{\bf k}^{ }$ with
$ \Psi_{\bf k}^\dag = \left[
e^{-ik_y d_1} c_{1,{\bf k}}^\dag,~ c_{2,{\bf k}}^\dag,~
e^{-ik_y d_1} c_{3,{\bf k}}^\dag,~ c_{4,{\bf k}}^\dag
\right]$,
\begin{equation}
\label{eqn:hamiltonian_k}
\!\!\!\hat{H}_{\bf k} \!=\!\!
\left[
\begin{array}{cccc}
\!\! \mu_c \!\! & \!\! 2t_{zz}\!\cos\frac{k_x a}{2}  \!\!
& \!\! 0 \!\! & \!\! t_{ac} e^{ik_y b/2} \!\!  \\
\!\! 2t_{zz}\!\cos\frac{k_x a}{2} \!\! & \!\! \mu_c \!\! 
& \!\! t_{ac}e^{-ik_y b/2} \!\! & \!\! 0 \!\! \\
\!\! 0 \!\! &  \!\! t_{ac} e^{ik_y b/2} \!\! 
& \!\! -\mu_c \!\!  & \!\! 2t_{zz}\!\cos\frac{k_x a}{2} \!\! \\   
\!\! t_{ac} e^{-ik_y b/2} \!\!  & \!\! 0 \!\! 
& \!\! 2t_{zz}\!\cos\frac{k_x a}{2} \!\! & \!\! -\mu_c  \!\!
\end{array}
\right]\!.
\end{equation}
Here, $c_{i,{\bf k}}^\dag (c_{i,{\bf k}}^{ })$ is the electron
creation (annihilation) operator of the orbital at the
$i$th P atom in the unit cell with the momentum ${\bf k}$,
and $a$ and $b$ are the unit-cell lengths
along the zigzag ($x$) and armchair ($y$) directions, respectively.
$d_1$ is the $y$-coordinate difference of the first and second atoms.
We also introduce a puckering potential energy, $\mu_c$, which
can be controlled by an electric field perpendicular to the BP plane.

BP has a glide-reflection (GR) symmetry with respect to the plane
at the middle of the zigzag chain
as shown by the red dotted plane in Fig.~\ref{fig:BPbiStruct}.
This symmetry exists regardless of the number of layers
and even with perpendicular electric fields that generate
nonzero intra- and interlayer potential differences.
As the GR operator commutes with Hamiltonian,
it is the Dirac-cone protecting nonsymmorphic 
symmetry \cite{Young2015} of gap-closed mono- and multilayer BP.

Using the GR symmetry, the Hamiltonian~(\ref{eqn:hamiltonian_k})
is block-diagonalized by a pseudospin
representation [see Supplemental Material]:
\begin{eqnarray}
\nonumber
\hat{H}_{{\bf k},1} &= &(D+2\!~t_{zz}\cos\frac{k_x a}{2})
\!~\hat{\sigma}_x 
+t_{ac}\sin\frac{k_y b}{2}\!~\hat{\sigma}_y, \\
\hat{H}_{{\bf k},2} &=& (D-2\!~t_{zz}\cos\frac{k_x a}{2})
\!~\hat{\sigma}_x 
-t_{ac}\sin\frac{k_y b}{2}\!~\hat{\sigma}_y,
\label{eqn:BP_eff_hamil_mono}
\end{eqnarray}
with $D = \sqrt{\mu_c^2+t_{ac}^2\cos^2(k_y b/2)}$.
Here, $\hat{\sigma}_x$ and $\hat{\sigma}_y$ in $\hat{H}_{{\bf k},i}$ 
are Pauli matrices with respect to the pseudospinors.
Energy bands from $\hat{H}_{{\bf k},1}$
and $\hat{H}_{{\bf k},2}$ are
$E_{{\bf k},1,\pm}\! =\! \pm \sqrt{\{D\!+\!2t_{zz}
\cos(k_x a/2)\}^2\!+\!t_{ac}^2\sin^2(k_y b/2)}$
and $E_{{\bf k},2,\pm} \!= \!\pm \sqrt{\{D\!-\!2t_{zz}
\cos(k_x a/2)\}^2\!+\!t_{ac}^2\sin^2(k_y b/2)}$,
respectively. Because $D \ge 0$ and $t_{zz} < 0$, the energy gap is 
determined by $\hat{H}_{{\bf k},1}$.

\begin{figure}
\includegraphics[width=3.8cm]{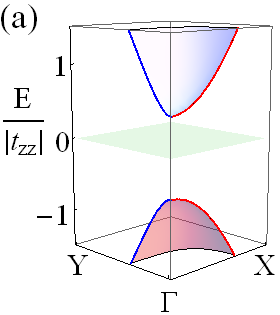}
\includegraphics[width=3.8cm]{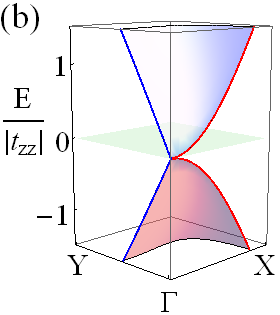}
\includegraphics[width=3.8cm]{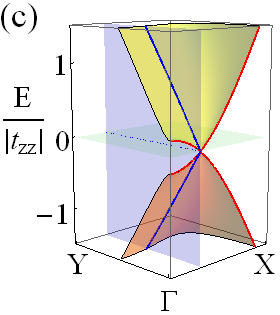}
\includegraphics[width=3.8cm]{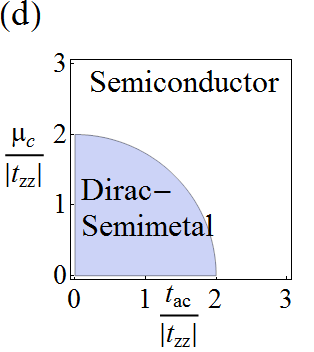}
\caption{
\label{fig:MonoTBBand}
(Color online) Tight-binding band structures near $\Gamma$ in monolayer BP for
(a) pristine, (b) $t_{ac}/|t_{zz}| = 2.0$,
and (c) $t_{ac}/|t_{zz}| = 1.8$.
(d) Phase diagram of monolayer BP.
For overall band structures along high-symmetry lines, see Fig.~S1
in Supplemental Material.
}
\end{figure}

Figure~\ref{fig:MonoTBBand} shows our tight-binding band structures
for different parameters. As seen in Fig.~\ref{fig:MonoTBBand}(a),
pristine monolayer BP has a direct band gap 
at $\Gamma$ in the Brillouin zone (BZ). 
Near $\Gamma$, the Hamiltonian $\hat{H}_{{\bf k},1}$ 
in \eqn{eqn:BP_eff_hamil_mono} can be approximated as
\begin{equation}
 \hat{H}_{{\bf k},1} \cong
\Big(\frac{1}{2}E_g\!+\!\frac{\hbar^2 k_x^2}{2 m_x}
+\frac{\hbar^2k_y^2}{2 m_y}-\frac{\hbar^2 v_y^2 k_y^2}{E_g}
\Big)\hat{\sigma}_x \!+ \hbar v_y k_y \hat{\sigma}_y, 
\label{eqn:near_gamma}
\end{equation}
where the energy gap
$E_g = 2\sqrt{\mu_c^2+t_{ac}^2}+4t_{zz}$, 
the zigzag effective mass $m_x=-2\hbar^2/(t_{zz}a^2)$,
the armchair effective mass 
$m_y = -\hbar^2E_g\sqrt{\mu_c^2+t_{ac}^2}/(t_{zz}t_{ac}^2b^2)$,
and the armchair velocity $v_y=t_{ac}b/(2\hbar)$.
The expression~(\ref{eqn:near_gamma}) is valid for positive or 
negative $E_g$. For zero $E_g$, all $k_y^2\hat{\sigma}_x$ terms
should be neglected together with $E_g$.
The band dispersions in $\Gamma$Y,
that is, the armchair direction from $\Gamma$, are
\begin{eqnarray}
E_{{\bf k},1,\pm} 
\simeq
\pm \sqrt{E_g^2/4
+|t_{zz}|t_{ac}^2 b^2 k_y^2/(2\sqrt{\mu_c^2+t_{ac}^2})},
\label{eqn:hyperbolic_disp}
\end{eqnarray}
which are hyperbola whose asymptotic lines are
$E = \pm \sqrt{t_{ac}^2b^2|t_{zz}|/(2\sqrt{\mu_c^2+t_{ac}^2})}~k_y$.

Monolayer BP is a honeycomb lattice in the sense that every P atom has 
three nearest neighbors. In the honeycomb lattice, an energy gap exists 
if the absolute values of the hopping energies to the three nearest 
neighbors cannot
form a triangle \cite{Hasegawa2006,Pereira2009,Kim2012a,Hou2015}. 
In the case of pristine BP, $\mu_c = 0$ and there is a finite energy gap of
$E_g = 1.514$~eV~$>0$. This nonzero band gap is consistent with the 
triangular criterion, that is, monolayer BP is semiconducting because 
the hopping energy along the armchair direction is greater than twice 
the absolute value of the hopping energy along the zigzag direction. 
If $t_{ac}$ decreases or $|t_{zz}|$ increases, $E_g$ will decrease. 
Thus, a way to modify the energy gap in BP layers is to apply pressure 
or strain which can change bond angles and bond lengths and thereby
the intralayer hopping energies.

The above triangular criterion is modified when an external electric 
field is applied. We define a gap-closing parameter including the
puckering potential energy $\mu_c$,
\beq
\beta_1 \equiv \left|\frac{t_{ac}}{2t_{zz}}\right|
\sqrt{1+\frac{\mu_c^2}{t_{ac}^2}}.
\label{eqn:betamono}
\eeq
Then, monolayer BP has a finite energy gap when $\beta_1> 1$, 
as in the case of the pristine monolayer [Fig.~\ref{fig:MonoTBBand}(a)],
and the gap vanishes when $\beta_1\le 1$.
At the moment of the gap closing ($\beta_1=1$),
the Hamiltonian $\hat{H}_{{\bf k},1}$ becomes highly anisotropic 
Dirac Hamiltonian 
\begin{eqnarray}
\hat{H}_{{\bf k},1} & \cong &
\frac{\hbar^2 k_x^2}{2 m_x} \hat{\sigma}_x 
+\hbar v_y k_y \hat{\sigma}_y,
\end{eqnarray}
which results in a linear dispersion in the armchair direction 
($\Gamma$Y) and quadratic one in the zigzag direction ($\Gamma$X) 
[Fig.~\ref{fig:MonoTBBand}(b)].
This corresponds to the single merged Dirac point of
the two separate Dirac points \cite{Montambaux2009}.

When $\beta_1 < 1$, monolayer BP is metallic, with the valence
and conduction bands touching each other 
at finite momenta $(\pm k_D,0)$ in the $\Gamma$X line 
[Fig.~\ref{fig:MonoTBBand}(c)].
From the condition that $E_{{\bf k},1,+} = E_{{\bf k},1,-}$, we obtain
\beq
k_D = (2/a)\cos^{-1}\!\beta_1.
\label{eqn:dirac_momentum}
\eeq
Near the crossing point, $(k_D,0)$,
the Hamiltonian $\hat{H}_{{\bf k},1}$ can be written as
a massless Dirac particle,
\begin{eqnarray}
\hat{H}_{{\bf k},1} & \cong & 
\hbar v_x (k_x-k_D)
\hat{\sigma}_x + v_y k_y \hat{\sigma}_y,
\label{eqn:eff_hamil_dirac}
\end{eqnarray}
where 
$v_x = -(t_{zz}a/\hbar)\sin(k_D a/2)=-(t_{zz}a/\hbar)\sqrt{1-\beta_1^2}$.
Thus, $(\pm k_D,0)$ are Dirac points with anisotropic linear bands
$E_{\pm}(\bk) =  \pm\hbar\sqrt{v_x^2 (k_x- k_D)^2 +v_y^2 k_y^2}$.
Perpendicular electric field can generate $\mu_c$, which re-opens 
the gap if it is large enough to make $\beta_1 >  1$.
Now the phase diagram is constructed according to \eqn{eqn:betamono}
in the parameter space of $(t_{ac},\mu_c)/|t_{zz}|$ 
[Fig.~\ref{fig:MonoTBBand}(d)].

The pseudospin representations of the conduction- and valence-band states are 
$\frac{1}{\sqrt{2}}(1, e^{i \phi})$ and 
$\frac{1}{\sqrt{2}}(1, -e^{ i \phi})$, respectively, when the 
Hamiltonian $\hat{H}_{ {\bf k},1}$ in \eqn{eqn:BP_eff_hamil_mono} is 
expressed as 
$\hat{H}_{{\bf k},1} = E_{{\bf k},1,+}(\cos\phi\!~\hat{\sigma}_x
+\sin\phi\!~\hat{\sigma}_y)$. Then, the expectation value of the pseudospin 
vector, $\vec{\sigma}$, for a conduction-band state is 
$\langle\vec{\sigma}\rangle
=\langle(\hat\sigma_x,\hat\sigma_y,\hat\sigma_z)\rangle
=(\cos\phi,\sin\phi,0)$, which lies in the $xy$-plane
in our representation. Figure~\ref{fig:pseudospin} shows the 
calculated $\langle \vec{\sigma}\rangle$ of the conduction and valence bands.
We find that the $y$-component $\langle\hat\sigma_y\rangle$ of each band 
always changes the sign by the reflection across $x$-axis, that is 
\beq
\langle \hat\sigma_y(k_x,k_y)\rangle 
= -\langle \hat\sigma_y(k_x,-k_y)\rangle 
\eeq
in the semiconducting [Fig.~\ref{fig:pseudospin}(a)] as well as 
Dirac-semimetallic phases [Figs.~\ref{fig:pseudospin}(b) and (c)].
This sign-reversal reduces the back scattering of charge carriers
moving along the armchair direction, enhancing their mobility even at the 
semiconducting phase, unless impurities or defects produce very abrupt 
potential. After the gap closing, the pseudospin becomes chiral around 
the Dirac points [Fig.~\ref{fig:pseudospin}(c)], enhancing the
mobility in all directions.

\begin{figure}
\includegraphics[width=8.5cm]{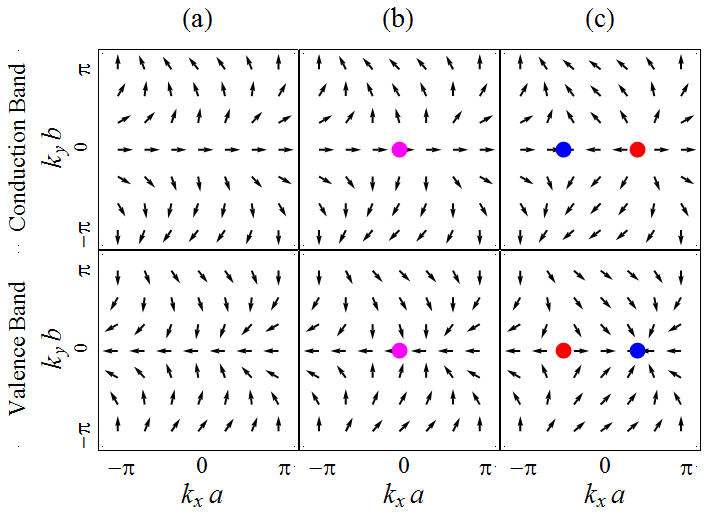}
\caption{
(Color online)
Pseudospin expectation values of conduction and valence bands
in BZ for (a) the semiconducting, (b) the highly anisotropic 
Dirac semimetallic, and (c) the Dirac semimetallic phase.
Colored dots in (b) and (c) are Dirac points.
\label{fig:pseudospin}
}
\end{figure}

BP has weak spin-orbit coupling (SOC) and the first-principles calculation
showed that SOC takes place as $\hat{H}_{\mbox{\scriptsize SOC}} =
	\lambda_{\mbox{\scriptsize SOC}}\hat S_x\hat{\sigma}_y$
in the low-energy effective Hamiltonian \cite{Baik2015}, where
$\lambda_{SOC}$ is a constant and $\hat S_x$ is the $x$-component
of the real spin. Since this term does not break the GR symmetry, 
the Dirac point in semimetallic BP layers
is still protected and only shifted in the $k_y$-direction.
To open a band gap at the Dirac point, we need a mass term,
which adds a $\hat{\sigma}_z$ term to \eqn{eqn:eff_hamil_dirac}.
If the mass term is simply a constant times $\hat{\sigma}_z$,
BP layers become a trivial insulator. However, if one can introduce, 
for example, a momentum-dependent mass term like
\beq
H_{\mbox{\scriptsize M}} =
\lambda_{\mbox{\scriptsize M}}\sin (k_x a)\hat{\sigma}_z S_z,
\label{eqn:soc_int}
\eeq
it generates a topologically nontrivial energy gap at the Dirac 
point \cite{Haldane1988}. Realization and control of such perturbation 
in BP is of great interest because it will enable 
a topologically trivial-to-nontrivial phase transition.

\begin{figure}
\includegraphics[width=3.8cm]{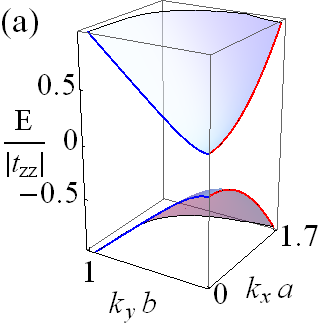}
\includegraphics[width=3.8cm]{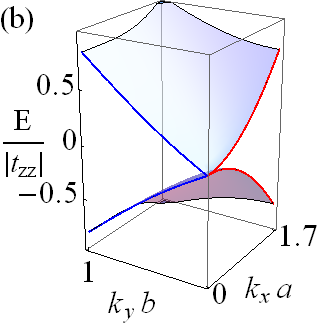}
\includegraphics[width=3.8cm]{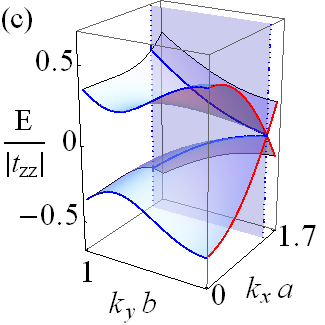}
\includegraphics[width=3.8cm]{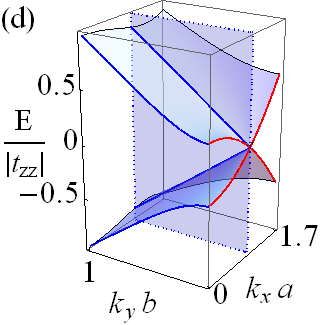}
\includegraphics[width=3.8cm]{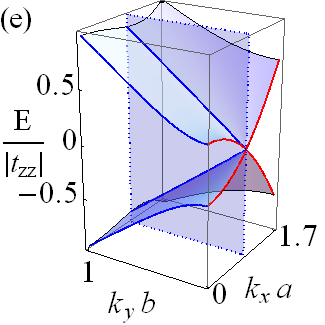}
\includegraphics[width=3.8cm]{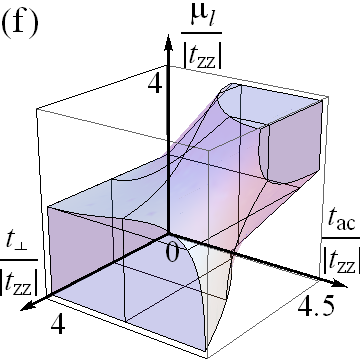} 
\caption{
(Color online)
Tight-binding band structures near $\Gamma$ in bilayer BP for
(a) pristine, (b) $\mu_l/|t_{zz}|=0.41$, (c) $\mu_l/|t_{zz}|=1.0$, 
(d) $t_{ac}/|t_{zz}|=2.05$, and (e) $t_\perp/|t_{zz}|=0.95$.
(f) Phase diagram of bilayer, where the dark region is the Dirac semimetal.
For overal band structures along high-symmetry lines, see Fig.~S2
in Supplemental Material.
\label{fig:BPdispEvol} }
\end{figure}

Now we consider bilayer BP. We generalize the operator 
$c_{i,{\bf k}}$ to $c_{i,{\bf k},j}$, where $j = 1,2$ indicates 
the first and second layer in the bilayer, and 
similarly $\Psi_{\bf k}$ to $\Psi_{{\bf k},j}$.
Then, the Hamiltonian of the bilayer is
\begin{eqnarray}
\nonumber
H  &= &\sum_{\bf k} \Big\{
\Psi_{{\bf k},1}^\dag \big(\hat{H}_{\bf k}+\mu_l\big)
\Psi_{{\bf k},1}^{ }
+\Psi_{{\bf k},2}^\dag \big(\hat{H}_{\bf k}-\mu_l\big)
\Psi_{{\bf k},2}^{ } 
\\ && ~~~
+\Psi_{{\bf k},1}^\dag \hat{V}_{\bf k}\Psi_{{\bf k},2}^{ }
+\Psi_{{\bf k},2}^\dag \hat{V}^\dagger_{\bf k}\Psi_{{\bf k},1}^{ }
\Big\},
\end{eqnarray}
where $\hat{H}_{\bf k}$ is the intralayer Hamiltonian 
(\ref{eqn:hamiltonian_k}), $\mu_l$ is half of the difference in the 
average potentials of the two layers, and $\hat{V}_{\bf k}$ is due to 
the interlayer hopping. We assume that the first layer is placed 
on top of the second layer so that the third and fourth P atoms 
of the first layer are the nearest neighbors of the first and second P 
atoms of the second layer, as shown in 
Fig.~\ref{fig:BPbiStruct}. Then, $\hat{V}_{\bf k}$ is a $4\times 4$
matrix 
$\left(\begin{array}{cc}
	\hat{0} & \hat{0} \\
	\hat{\nu}_{\bf k} & \hat{0}
\end{array}\right)$ with
$\hat{\nu}_{\bf k} = 2t_\perp\cos\frac{k_x a}{2}(
\cos\frac{k_yb}{2}\hat{\sigma}_x
- \sin\frac{k_y b}{2}\hat{\sigma}_y)$
and $\hat{0}$ is $2\times 2$ zero matrix.

Figures~\ref{fig:BPdispEvol}(a)-(e) show our tight-binding band 
structures of bilayer BP. 
Due to the GR symmetry of the lattice, the band-gap closing
occurs at a point in the $\Gamma$X line in BZ \cite{VanMiert2016}.
For $k_y = 0$, one can exploit the GR symmetry and obtain
the lowest conduction-band energy $E_c$ and
the highest valence-band energy $E_v$ in the
$\Gamma$X as
\begin{equation}
	E_{c,v}(k_x) = \pm 2t_{zz}\cos(k_xa/2)
\pm\sqrt{P_{k_x}- 2\sqrt{Q_{k_x}}}
\end{equation}
where $+$ and $-$ are for conduction band and valenece band,
respectively, 
$P_{k_x} = 2 t_\perp^2\cos^2(k_xa/2)
+t_{ac}^2+\mu_l^2+\mu_c^2$, and $Q_{k_x} =
t_{ac}^2t_\perp^2\cos^2(k_xa/2)+t_{ac}^2\mu_l^2
+\{t_\perp^2\cos^2(k_xa/2)-\mu_l\mu_c\}^2$.
Then the Dirac point $k_D$ is determined by the condition $E_c(k_D) = E_v(k_D)$.
For the simplicity, we neglect $\mu_c$,
since $\mu_c$ has a smaller value than $\mu_l$ at an applied electric field.
When $\mu_c = 0$, we define the gap-closing parameter 
$\beta_2$ of bilayer BP as
\begin{equation}
\beta_2  =  \sqrt{
\frac{2t_{zz}^2(t_{ac}^2+\mu_l^2)-t_\perp^2\mu_l^2-\sqrt{F}}
{8t_{zz}^2(t_{zz}^2-t_\perp^2)}}
\label{eqn:betabi}
\end{equation}
with $F=4t_{zz}^2t_{ac}^2(t_\perp^2t_{ac}^2
+4t_{zz}^2\mu_l^2-3t_\perp^2\mu_l^2)+t_\perp^4\mu_l^4$.
When $\beta_2 > 1$, bilayer BP is semiconducting with a direct band gap
at $\Gamma$. When $\beta_2 < 1$, bilayer BP is metallic with Dirac points
at $(\pm k_D,0)$ given by
\begin{equation}
k_D  =  (2/a)\cos^{-1}\!\beta_2.
\end{equation}
In the case of pristine bilayer, $\beta_2 > 1$ so that it is 
semiconducting, as shown in Fig.~\ref{fig:BPdispEvol}(a). Here, due to 
the interlayer coupling $t_\perp$, the gap is smaller than in the monolayer.

The interlayer potential energy difference $\mu_l$, which can be generated 
by applied electric field, decreases the energy gap of bilayer BP. 
As shown in Figs.~\ref{fig:BPdispEvol}(b) and (c), the increase of 
$\mu_l$ can close the band gap and induce the Dirac-semimetal phase.
This is in contrast with the puckering potential energy $\mu_c$ 
that acts against the gap closing 
although $\mu_c$ is also generated by an electric field.

Figures~\ref{fig:BPdispEvol}(d) and (e) show the band-gap closing
by reducing the armchair hopping and by increasing the interlayer
hopping, respectively. Compared with these bands, the field-induced band 
structure in Fig.~\ref{fig:BPdispEvol}(c) displays non-monotonous 
dispersions of the conduction and valence bands in the $\Gamma$Y line, 
where a minimum of their band-energy difference occurs away from $\Gamma$.

In bilayer BP, the interlayer coupling $t_\perp$ decreases $\beta_2$ 
of \eqn{eqn:betabi}, making the Dirac point appear more easily in bilayer 
than monolayer. Since the GR symmetry exists in multilayer BP, 
$t_\perp$ does not mix different pseudospins of each layer, but it widens 
each band, resulting in a smaller band gap in a thicker BP layer. 
When the band gap is closed, the band dispersions from $\Gamma$ to Y
in Fig.~\ref{fig:BPdispEvol}(e) are monotonous similarly 
to the case of the armchair-hopping reduction [Fig.~\ref{fig:BPdispEvol}(d)], 
in contrast with the non-monotonous behavior in Fig.~\ref{fig:BPdispEvol}(c), 
as mentioned above. Applied pressure can contribute to the band-gap 
reduction and closing since it enhances the interlayer coupling by 
decreasing the interlayer distance.

From \eqn{eqn:betabi}, we obtained the phase diagram of Dirac semimetal 
for bilayer BP in the three-dimensional parameter space of 
$(t_{ac},\mu_l,t_\perp)/|t_{zz}|$, as shown in Fig.~\ref{fig:BPdispEvol}(f).
Here the parameters $t_{ac}$, $\mu_l$, and $t_{\perp}$ are changeable 
by strain, pressure, and electric fields, as discussed above.
We find that the region for the Dirac semimetal phase is continuously 
connected [Fig.~\ref{fig:BPdispEvol}(f)], indicating that the 
Dirac-semimetal phase is topologically identical regardless of
the gap-closing method.

In summary, we obtained a unified picture on the Dirac semimetal phase 
of BP layers. Phase diagrams are obtained for mono- and bilayer BP 
in the parameter space of the Hamiltonian. Phase-controlling parameters 
are identified to be (i)~the anisotropy of the intralayer hopping,
which can be changed by pressure or strain, (ii)~the interlayer hopping, 
which can be enhanced by pressure, and (iii)~the interlayer potential 
energy difference, which can be generated by electric fields. Pseudospins, 
originating from the glide-reflection symmetry of BP, exist even in 
the semiconducting phase and assure the Dirac-semimetal phase when the 
band gap is closed. The Dirac-semimetal phase in BP layers is singly 
connected in the parameter space, indicating it is of the same kind 
regardless of the gap-closing method. Our findings can be generalized 
to the Dirac-semimetal phase in anisotropic layered materials and can 
play a guiding role in search for a new class of topological materials 
and devices. 

\begin{acknowledgments}
This work was supported by the NRF of Korea (Grant No. 2011-0018306). 
\end{acknowledgments}

\clearpage


\title{Supplemental Material: \\
Dirac-semimetal phase diagram of two-dimensional black
phosphorus}

\author{Hyeonjin Doh}
\email[]{clotho72@yonsei.ac.kr}
\affiliation{Department of Physics and
Center for Computational Studies of Advanced Electron Material Properties,
Yonsei University, Seoul 03722, Korea}

\author{Hyoung Joon Choi}
\email[]{h.j.choi@yonsei.ac.kr}
\affiliation{Department of Physics and
Center for Computational Studies of Advanced Electron Material Properties,
Yonsei University, Seoul 03722, Korea}


\date{\today}


\maketitle


\begin{figure*}
\includegraphics[width=7.4cm]{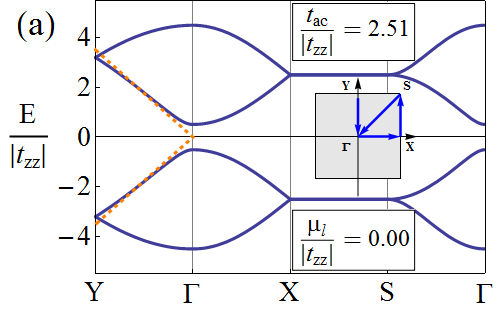}
\includegraphics[width=7.4cm]{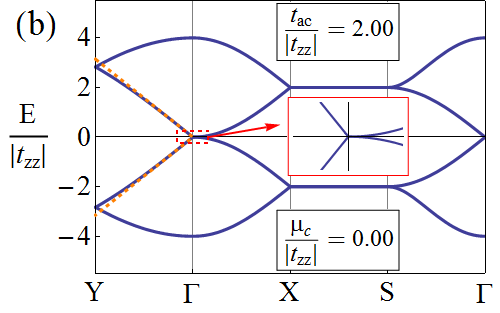}
\includegraphics[width=7.4cm]{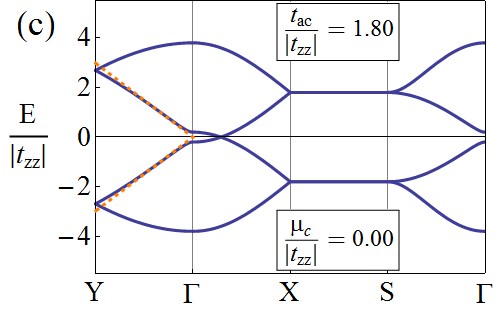}
\includegraphics[width=7.4cm]{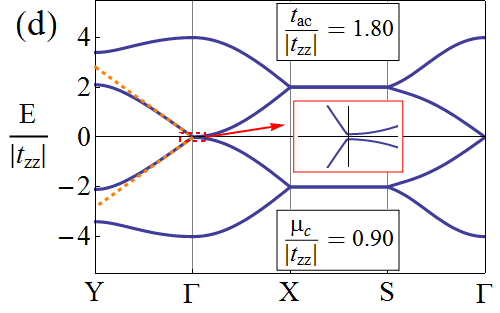}
\caption{
Tight-binding band structures along high-symmetry lines in 
monolayer BP for (a) the pristine case, (b) $t_{ac}$ reduced to $2|t_{zz}|$, 
(c) $t_{ac}$ reduced to $1.8|t_{zz}|$, and (d) $t_{ac}$ reduced to $1.8|t_{zz}|$
together with $\mu_c$ set to $0.9|t_{zz}|$. 
In (b)-(d), orange dashed lines are asymptotic lines.
The inset in (a) shows the high-symmetry lines in the two-dimensional Brillouin zone.
The insets in (b) and (d) enlarge band dispersions near $\Gamma$.}
\end{figure*}
\begin{figure*}
\includegraphics[width=7.3cm]{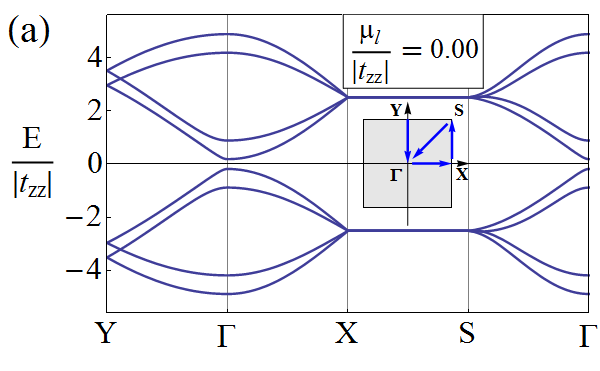}
\includegraphics[width=7.3cm]{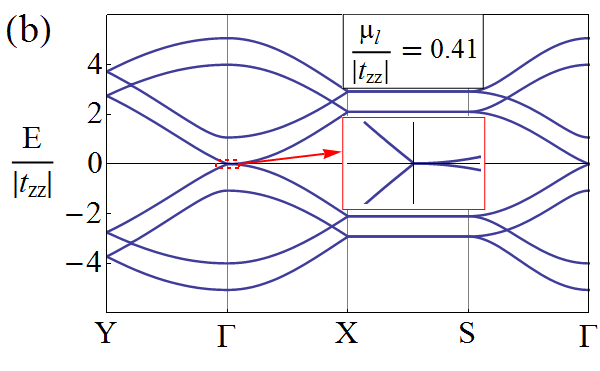}
\includegraphics[width=7.3cm]{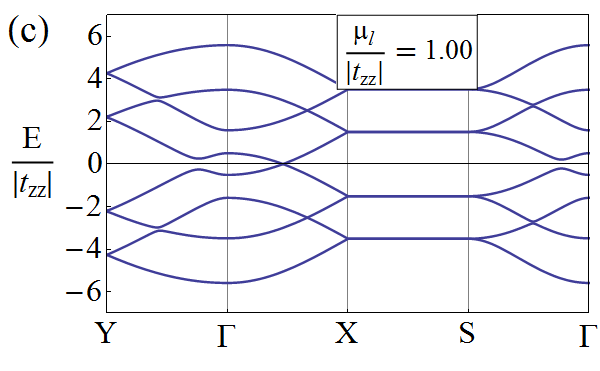}
\includegraphics[width=7.3cm]{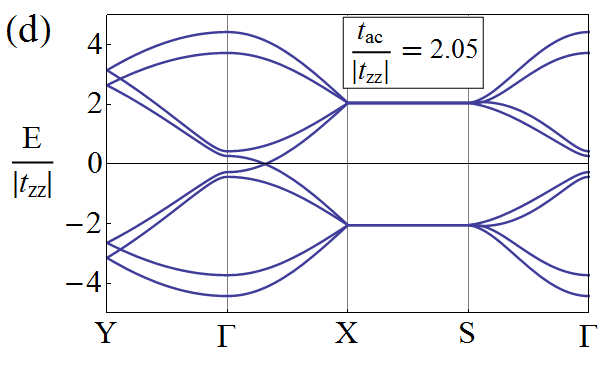}
\includegraphics[width=7.3cm]{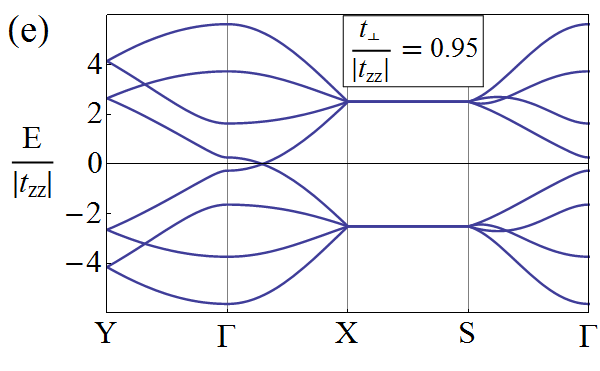}
\includegraphics[width=7.3cm]{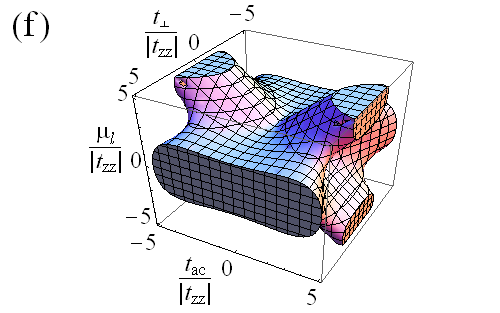}
\caption{
Tight-binding band structures along high-symmetry lines in bilayer BP for
(a) the pristine case, (b) $\mu_l$ set to $0.41|t_{zz}|$,
(c) $\mu_l$ set to $|t_{zz}|$, (d) $t_{ac}$ reduced to $2.05|t_{zz}|$,
and (e) $t_{\perp}$ increased to $0.95|t_{zz}|$.
The inset in (a) shows the high-symmetry lines in the two-dimensional Brillouin zone.
The inset in (b) enlarges the band dispersions near $\Gamma$.
(f) Region of the Dirac-semimetal phase in the parameter space
of $(t_{ac},\mu_l,t_\perp)/|t_{zz}|$. In (f), the parameter space is larger than
that in Fig.~4(f) in the main text which covers $t_{ac} \geq 0$, $\mu_l \geq 0$,
and $t_{\perp} \geq 0$ only.
}
\end{figure*}


\begin{center}
{\bf Derivation of Dirac Hamiltonian for\\
monolayer black phosphorus}
\end{center}

Our tight-binding Hamiltonian for the monolayer BP is expressed with local bases as
\beqa
\label{eqn:hamiltonian_real}
H\!=\! \mbox{$\sum_{\bf R}$}&& \big\{
t_{zz}\big(
 c_{1,{\bf R}}^\dag c_{2,{\bf R}}^{}
+c_{1,{\bf R}}^\dag c_{2,{\bf R}+a{\bf \hat{x}}}^{} \nonumber \\
&&~~~~~~~+c_{3,{\bf R}}^\dag c_{4,{\bf R}}^{}
+c_{3,{\bf R}}^\dag c_{4,{\bf R}-a{\bf \hat{x}}}^{}
+\mbox{h.c} \big)   \nonumber \\ 
&&
+t_{ac}\big(
 c_{2,\bf{R}}^\dag c_{3,\bf{R}}^{}
+c_{1,{\bf R}+b{\bf \hat{y}}}^\dag c_{4,{\bf R}}^{}
+\mbox{h.c} \big) 
\\ &&
+\mu_c\big(
 c_{1,\bf{R}}^\dag c_{1,\bf{R}}^{}
\!+\!c_{2,\bf{R}}^\dag c_{2,\bf{R}}^{}
\!-\!c_{3,\bf{R}}^\dag c_{3,\bf{R}}^{}
\!-\!c_{4,\bf{R}}^\dag c_{4,\bf{R}}^{}
\big)\!
\big\}.
\nonumber
\eeqa
Here, $c_{i,{\bf R}}^\dag(c_{i,{\bf R}}^{})$ is the electron
creation (annihilation) opearator of the orbital at the $i$th 
P atom in the unit cell at ${\bf R}$, and $a$ and $b$ are the
unit-cell lengths along the zigzag and armchair directions,
respectively. 
Here we set the $x$- and $y$-axis along zigzag and armchair
direction, respectively.
To obtain electric band structure,
we define
$c_{i,{\bf k}}=\frac{1}{\sqrt{N}}\sum_{\bf R} c_{i,{\bf R}}
e^{i{\bf k}\cdot({\bf R}+{\bf r}_i)}$,
where
${\bf r}_i$ is the position of the $i$th basis
within the unit cell
and 
N is the number of ${\bf k}$ vectors.
By Fourier transformation of
$H = \sum_{\bf k} \Psi_{\bf k}^\dag \hat{H}_{\bf k}\Psi_{\bf k}$
with
$\Psi_{\bf k}^\dag = \left[
c_{1,{\bf k}}^\dag e^{-ik_yd_1},
c_{2,{\bf k}}^\dag,
c_{3,{\bf k}}^\dag e^{-ik_yd_1},
c_{4,{\bf k}}^\dag
\right]$, we obtain the momentum representation $\hat{H}_{\bf k}$
of \eqn{eqn:hamiltonian_real}
as
\beq
\hat{H}_{\bf k}\! =\!\!
\left[\!
\begin{array}{cccc}
\mu_c \!\!&\!\! 2t_{zz}\!\cos\!\frac{k_x a}{2} \!\!&\!\! 0 \!\!&\!\! t_{ac}e^{i k_y b/2} \\
2t_{zz}\!\cos\!\frac{k_x a}{2} \!\!&\!\! \mu_c \!\!&\!\! t_{ac}e^{-i k_y b/2} \!\!&\!\! 0 \\
0 \!\!&\!\! t_{ac}e^{i k_y b/2} \!\!&\!\! -\mu_c \!\!&\!\! 2t_{zz}\!\cos\!\frac{k_x a}{2} \\
t_{ac}e^{-i k_y b/2} \!\!&\!\! 0 \!\!&\!\! 2t_{zz}\!\cos\!\frac{k_x a}{2} \!\!&\!\! -\mu_c 
\end{array} \!\right]\!\!,\!\!\!
\label{eqn:hamiltonian_k}
\eeq
which is Eq.~(1) of the main text.

To find a necessary condition for a pseudospin basis, we suppose
the low-energy
effective Hamiltonian of \eqn{eqn:hamiltonian_k} can be written
as a Dirac Hamiltonian with
a linear dispersion in $y$-direction
at $k_y=0$. Then, it is written as  
\beq
\hat{H}_{\bf k}^{eff} \simeq \cdots +
v_y k_y \hat{\sigma}_y+\cdots,
\eeq
in terms of a pseudospin basis.
Here $\hat{\sigma}_y$ is the Pauli matrix whose basis represents
a pseudospin in the BP system.
If a symmetry operation preserves the Dirac Hamiltonian and transforms $k_y$ to $-k_y$, the symmetry operation should transform the pseudospin $S_y$ to $-S_y$.
For monolayer BP, 
the glide-refleciton (GR) operator $T_{GR}$,
associated with the GR symmetry plane in Fig.~1 of the main text,
can be written as
\beq
T_{GR}= e^{-i p_y \frac{d_1}{2}}
e^{i X \frac{\pi}{2}}
e^{i p_y \frac{d_1}{2}}
e^{i p_x\frac{a}{2}},
\label{eqn:glide-reflection}
\eeq
where $X$, $p_x$, and $p_y$ operators are
\beqn
X &\equiv& \sum_{{\bf k}\nu\nu'}
\eta_{\nu\nu'}
c_{\nu,(k_x,-k_y)}^\dag
c_{\nu',(k_x,k_y)}^{},
\\  \nonumber
p_x&\equiv& \sum_{{\bf k}\nu} 
k_x c_{\nu,(k_x,k_y)}^\dag c_{\nu,(k_x,k_y)}^{ }, 
\\
p_y&\equiv& \sum_{{\bf k}\nu} 
k_y c_{\nu,(k_x,k_y)}^\dag c_{\nu,(k_x,k_y)}^{ },
\eeqn
respectively. Here $\eta_{\nu\nu'} = 1$ 
for $(\nu,\nu') = (1,2)$, $(2,1)$, $(3,4)$, and $(4,3)$, 
and it is zero otherwise. The Hamiltonian 
\eqn{eqn:hamiltonian_k} is symmetric under the
glide-reflection operation and contains odd function of
$k_y$ in its matrix elements. Therefore, 
for the Hamiltonian to be written as a Dirac Hamiltonian in 
a pseudospin representation,
the pseudospin operator should have the following symmetric
property under the glide-reflection operator:
\beqa
T_{GR}^{ } S_x T_{GR}^\dag &= &S_x, \nonumber \\
T_{GR}^{ } S_y T_{GR}^\dag  &= & -S_y.
\eeqa
Here,
$S_i \equiv \sum_{\sigma\sigma'\nu} \chi_{\sigma,\nu}^\dag
(\hat{\sigma}_i)_{\sigma\sigma'}
\chi_{\sigma',\nu}^{}$ is a pseudospin operator for $i=x,y,z$ and
$\chi_{\sigma,\nu}^{}$ is a pseudospinor for 
$\sigma=\uparrow$, $\downarrow$. Since we have
four bases in a unit cell, we have two orthogonal pseudospins
which are distinguished by the index $\nu$ = 1, 2.
These symmetry conditions correspond to the following symmetric operation
on the pseudospinors:
\beq
T_{GR} \chi_{\sigma,\nu}^{} 
T_{GR}^\dag= e^{i\theta}\chi_{-\sigma,\nu}^{},
\label{eqn:gr_spin_flip}
\eeq
where the minus sign on $\sigma$ means the opposite spin.
The constant phase factor $\theta$ is
independent of $\sigma$ and $\nu$.

Now we assume the pseudospin state can be written in terms of
the basis function in the unit cell,
\beqa
\nonumber 
\chi_{\sigma,\nu}&=&
a_{\sigma\nu,1}c_{1,{\bf k}}e^{ik_y d_1}
+a_{\sigma\nu,2}c_{2,{\bf k}} \\
&+&a_{\sigma\nu,3}c_{3,{\bf k}}e^{ik_y d_1}
+a_{\sigma\nu,4}c_{4,{\bf k}}.
\label{eqn:pseudospinor}
\eeqa
The basis functions in the unit cell are transformed 
by the glide-reflection operator in the following way:
\beqa
\nonumber
T_{GR}~c_{1,(k_x,k_y)}^\dag ~\! T_{GR}^\dag
&=& i e^{ik_xa/2}e^{ik_yd_1} ~\! c_{2,(k_x,-k_y)}^\dag,\\
\nonumber
T_{GR}~c_{2,(k_x,k_y)}^\dag ~\! T_{GR}^\dag
&=& i e^{ik_xa/2}e^{-ik_yd_1}~\!  c_{1,(k_x,-k_y)}^\dag,\\
\nonumber
T_{GR}~c_{3,(k_x,k_y)}^\dag ~\! T_{GR}^\dag
&=& i e^{ik_xa/2}e^{ik_yd_1}~\!  c_{4,(k_x,-k_y)}^\dag,\\
T_{GR}~c_{4,(k_x,k_y)}^\dag ~\! T_{GR}^\dag
&=& i e^{ik_xa/2}e^{-ik_yd_1}~\!  c_{3,(k_x,-k_y)}^\dag.
\label{eqn:gr_basis_tr}
\eeqa
Using \eqn{eqn:gr_spin_flip},
\eqn{eqn:pseudospinor},
\eqn{eqn:gr_basis_tr}, and an orthogonal condition
$\chi_{\sigma,\nu}^{}\chi_{\sigma',\nu'}^\dag
+\chi_{\sigma',\nu'}^\dag\chi_{\sigma,\nu}^{}
= \delta_{\sigma\sigma'}\delta_{\nu\nu'}$,
the pseudospin states can be written as
\beqa
\nonumber
\chi_{\uparrow,1}^{} &=&
\sin\theta_1 \sin\theta_2 ~\! c_{1,{\bf k}} e^{i k_yd_1}
+\cos\theta_1 \cos\theta_2 ~\! c_{2,{\bf k}}\\
&-& \nonumber
\sin\theta_1 \cos\theta_2 ~\! c_{3,{\bf k}} e^{i k_yd_1}
+\cos\theta_1 \sin\theta_2 ~\! c_{4,{\bf k}},\\
\nonumber
\chi_{\downarrow,1}^{} &=&
\cos\theta_1 \cos\theta_2 ~\! c_{1,{\bf k}} e^{i k_yd_1}
+\sin\theta_1 \sin\theta_2 ~\! c_{2,{\bf k}} \\
&+& \nonumber
\cos\theta_1 \sin\theta_2 ~\! c_{3,{\bf k}} e^{i k_yd_1}
-\sin\theta_1 \cos\theta_2 ~\! c_{4,{\bf k}}, \\
\nonumber
\chi_{\uparrow,2}^{} &=&
\cos\theta_1 \sin\theta_2 ~\! c_{1,{\bf k}} e^{i k_yd_1}
-\sin\theta_1 \cos\theta_2 ~\! c_{2,{\bf k}} \\
&-& \nonumber
\cos\theta_1 \cos\theta_2 ~\! c_{3,{\bf k}} e^{i k_yd_1}
-\sin\theta_1 \sin\theta_2 ~\! c_{4,{\bf k}}, \\
\nonumber
\chi_{\downarrow,2}^{} &=&
\sin\theta_1 \cos\theta_2 ~\! c_{1,{\bf k}} e^{i k_yd_1}
-\cos\theta_1 \sin\theta_2 ~\! c_{2,{\bf k}} \\
&+&
\sin\theta_1 \sin\theta_2 ~\! c_{3,{\bf k}} e^{i k_yd_1}
+\cos\theta_1 \cos\theta_2 ~\! c_{4,{\bf k}},
\eeqa
where $\theta_1$ and $\theta_2$ are real numbers which are
not determined yet.
Now we block-diagonalize the Hamiltonian
\eqn{eqn:hamiltonian_k}
into two $2\times 2$ matrices using the pseudospin states.
We determine $\theta_1$ and $\theta_2$ to satisfy 
\beq
\langle \chi_{\sigma,1}^{}|\hat{H}_{\bf k}|\chi_{\sigma',2}^{}\rangle=
\langle \chi_{\sigma,2}^{}|\hat{H}_{\bf k}|\chi_{\sigma',1}^{}\rangle=0
\eeq
for $\sigma = \uparrow$, $\downarrow$ and $\sigma' = \uparrow$, $\downarrow$. 
This condition
yields
\beqa
\nonumber
\tan (2\theta_1) & = & \mu_c / (t_{ac}\cos(k_yb/2)), \\
\theta_2&=&\pi/4.
\eeqa
Thus, the pseudospinors are related to
the four atomic orbitals in the unit cell as
\beqa
\nonumber
\chi_{\uparrow,1}^{} = (
&\sin\theta_{\bf k}&  c_{1,{\bf k}}e^{ik_yd_1}
+\cos\theta_{\bf k}  c_{2,{\bf k}} \\
\nonumber 
-&\sin\theta_{\bf k}&  c_{3,{\bf k}}e^{ik_yd_1}
+\cos\theta_{\bf k}  c_{4,{\bf k}}~\! )/\sqrt{2}, \\
\nonumber
\chi_{\downarrow,1}^{} =(
&\cos\theta_{\bf k}&  c_{1,{\bf k}}e^{ik_yd_1}
+\sin\theta_{\bf k}  c_{2,{\bf k}} \\
\nonumber 
+&\cos\theta_{\bf k}&  c_{3,{\bf k}}e^{ik_yd_1}
-\sin\theta_{\bf k}  c_{4,{\bf k}}~\! )/\sqrt{2}, \\
\nonumber
\chi_{\uparrow,2}^{} = (
&\cos\theta_{\bf k}&  c_{1,{\bf k}}e^{ik_yd_1}
-\sin\theta_{\bf k}  c_{2,{\bf k}} \\
\nonumber 
-&\cos\theta_{\bf k}&  c_{3,{\bf k}}e^{ik_yd_1}
-\sin\theta_{\bf k}  c_{4,{\bf k}}~\! )/\sqrt{2}, \\
\nonumber
\chi_{\downarrow,2}^{} =  (
&\sin\theta_{\bf k}&  c_{1,{\bf k}}e^{ik_yd_1}
-\cos\theta_{\bf k}  c_{2,{\bf k}} \\
+&\sin\theta_{\bf k}&  c_{3,{\bf k}}e^{ik_yd_1}
+\cos\theta_{\bf k}  c_{4,{\bf k}}~\! )/\sqrt{2}
\eeqa
with
$\sin(2\theta_{\bf k}) = \m_c/D$, 
$\cos(2\theta_{\bf k}) = (t_{ac}/D)\cos(k_yb/2)$,
and $D = \sqrt{\m_c^2+t_{ac}^2\cos^2(k_yb/2)}$.
We define pseudospinor vector $\hat{X}_{\bf k}^\dag=
[\chi_{\uparrow,1,{\bf k}}^\dag,
\chi_{\downarrow,1,{\bf k}}^\dag,
\chi_{\uparrow,2,{\bf k}}^\dag,
\chi_{\downarrow,2,{\bf k}}^\dag]$.
Then its relation to $\hat{\Psi}_{\bf k}$ can be expressed as
\beq
\hat{X}_{\bf k} = \hat{U}_{\bf k}\hat{\Psi}_{\bf k},
\eeq
using a unitary matrix
\beqa
\hat{U}_{\bf k} &=& \frac{1}{\sqrt{2}}\!\left[
\begin{array}{rrrr}
 \sin\theta_{\bf k} &  \cos\theta_{\bf k} &
-\sin\theta_{\bf k} &  \cos\theta_{\bf k} \\
 \cos\theta_{\bf k} &  \sin\theta_{\bf k} &
 \cos\theta_{\bf k} & -\sin\theta_{\bf k} \\
 \cos\theta_{\bf k} & -\sin\theta_{\bf k} &
-\cos\theta_{\bf k} & -\sin\theta_{\bf k} \\
 \sin\theta_{\bf k} & -\cos\theta_{\bf k} &
 \sin\theta_{\bf k} &  \cos\theta_{\bf k} 
\end{array}
\right]\!\!.~~~
\eeqa
These pseudospin bases
block-diagonalize the Hamiltonian~(\ref{eqn:hamiltonian_k})
to two $2\times 2$ blocks:
\beq
\hat{U}_{\bf k}^{} \hat{H}_{\bf k} \hat{U}_{\bf k}^\dag
\!=\!\!\left[
\begin{array}{cccc}
0 & \!\!  f^+_{\bf k}\! -\!i  g_{\bf k}  \!\!& 0 & 0 \\
f^+_{\bf k}\! +\!i  g_{\bf k} \!\! & 0 & 0 & 0 \\
0 & 0 & 0 & \!\!  f^-_{\bf k}\! +\!i  g_{\bf k} \\
0 & 0 &  \!\!  f^-_{\bf k}\! -\!i  g_{\bf k}  \!\!& 0
\end{array}
\right]
\eeq
with
\beqa
\nonumber
f^\pm_{\bf k} &= &\sqrt{\mu_c^2+t_{ac}^2\cos^2(k_yb/2)}
\pm 2t_{zz}\cos(k_x a/2), \\
\nonumber
g_{\bf k} &=& t_{ac}\sin(k_yb/2).
\eeqa
Now we have two Dirac-type Hamiltonians
\beqa
\nonumber
\hat{H}_{{\bf k},1} &=&
f^+_{\bf k}~\hat{\sigma}_x +g_{\bf k}~\hat{\sigma}_y,  \\
\hat{H}_{{\bf k},2} &=&
f^-_{\bf k}~\hat{\sigma}_x -g_{\bf k}~\hat{\sigma}_y,
\eeqa
which correspond to Eq.~(2) of the main text. These Dirac-type
Hamiltonians are valid in the semiconducting phase as well as the Dirac-semimetal
phase.

\begin{center}
{\bf Tight-binding band structures along high-symmetry lines\\}
\end{center}

Figures~S1 and S2 show overall electronic band structures along high-symmetry
lines in mono- and bilayer black phosphorus (BP) obtained from our tight-binding model described in the main text.
Figures~S1 and S2 correspond to Figs.~2 and 4 in the main text, respectively.

\end{document}